# Comment on 'Encoding many channels on the same frequency through radio vorticity: first experimental test'

**Michele Tamagnone**[1], **Christophe Craeye**[2,3] **and Julien Perruisseau-Carrier**[1,4]

[1] Adaptive MicroNanoWave Systems Group, LEMA/Nanolab, Ecole Polytechnique Fédérale de Lausanne (EPFL), 1015 Lausanne, Switzerland
[2] Université Catholique de Louvain (UCL), ICTEAM Institute. Bâtiment Stévin - 2, Place du Levant B-1348 Louvain-la-Neuve, Belgique
[3] E-mail : christophe.craeye@uclouvain.be
[4] E-mail: julien.perruisseau-carrier@epfl.ch

**Abstract**: We show that the public experiment held in Venice by F. Tamburini *et al* and reported in (2012 *New J. Phys.* **14** 033001) can be regarded as a particular implementation of Multiple-Input Multiple-Output (MIMO) communications, hence bringing no advantages with respect to known techniques. Moreover, we explain that the use of a 'vortex' mode (orbital angular momentum OAM $\ell = 1$) at one of the transmit antennas is not necessary to encode different channels since only different patterns –or similarly different pointing angles– of the transmit antennas are required. Finally, we identify why this MIMO transmission allowed decoding of two signals despite being line-of-sight. This is due to the large separation between the receiving antennas, which places the transmit antennas in the near-field Fresnel region of the receiving 'array'. This strongly limits the application of this technique in practice, since, for a fixed separation between receiving antennas, the detectable signal power from any additional vortex mode decays at least as $1/r^4$.

**Acknowledgements**: The authors would like to thank Ivano Adolfo Maio from Politecnico di Torino, Italy, for the fruitful discussion.

## 1. Introduction: description of the experiment

The experiment performed in Venice on June 24 2011 by Tamburini *et al*. and described in [1] aimed to show how radio vorticity could be used to encode several communication channels on the same frequency band. To this aim, two transmit (Tx) antennas were placed on the light house of San Giorgio Island, and two receive (Rx) antennas were placed on the balcony of Palazzo Ducale in Venice. The distance $r$ between the two 'arrays' was approximately 442 m. One of the two Tx antennas was a commercial Yagi-Uda antenna producing a classical directive radiation pattern interpreted as a $\ell = 0$ OAM (Orbital Angular Momentum) mode. The other Tx antenna was a modified reflector able to produce a pattern corresponding to an OAM mode with $\ell = 1$. Note that although the latter claim is not supported by antenna simulation or phase pattern measurements in [1], we will assume that this is correct. The two Rx antennas were also of Yagi-Uda type, and were separated by a distance $d$ of about 4.5 m. All antennas had horizontal polarization. By tuning the system and then *summing and subtracting* the signals received by the two Rx antennas, the team managed to reconstruct the two distinct signals transmitted by the two Tx antennas at the same time.

## 2. Analysis of the experiment

The aforementioned experimental setup can be considered as a point-to-point 2×2 Multiple Input Multiple Output (MIMO) antenna system. In fact, in general, a point-to-point $M \times N$ MIMO system is characterized by a Tx node comprising *M* Tx antennas and a Rx node with *N* Rx antennas (in this definition a radiating structure having two input ports, such as for instance a dual-polarized antenna, makes up *two* antennas). This definition is totally general and is thus valid independently of the type of antennas and so it also applies in [1]. It is also worth mentioning that other authors have recently discussed OAM modes from the perspective of MIMO communications [2], thereby supporting the approach followed here while providing different but complementary information. In particular [2] provided a general theoretical analysis of OAM modes at about the same time as [1], while the present comment obviously focuses on analyzing –notably intuitively– the experiment of [1]. Since Maxwell's equations are linear, the signals received by the Rx antennas will be linear combinations of the signals transmitted by the Tx antennas:

$$y = Hs \qquad (1)$$

where *y* is the vector of the received signals, *s* is the vector of transmitted signals and *H* is the matrix of coefficients (for simplicity, and without loss of generality, noise at the receiver is neglected here).

In analog MIMO systems, it is in general possible to decode the original transmitted signals (if $N \geq M$) by exploiting the fact that the linear combination coefficients are potentially different in each receiving antenna (assuming a deterministic knowledge of the coefficients). In general these differences are due to significant scattering, but can also arise when one of the arrays is located in (or close enough to) the near-field of the other array, which is the case of the Venice demonstration [1] or of an alternative similar experiment which we propose hereunder.

The detection in such a MIMO system can then be achieved by multiplying the received signal *y* by the pseudo-inverse of *H*. This is exactly what is done by the 'interferometer' of [1], since from the provided experiment description we have (*Y* and *V* being two system-dependent complex parameters):

$$H = \begin{pmatrix} Y & V \\ Y & -V \end{pmatrix} \rightarrow H^{-1} = \frac{1}{2YV}\begin{pmatrix} V & V \\ Y & -Y \end{pmatrix} \rightarrow \hat{s} = \frac{1}{2YV}\begin{pmatrix} V & V \\ Y & -Y \end{pmatrix} y \qquad (2)$$

In other words, in [1] the OAM ℓ = 0 signal is reconstructed (first component of $\hat{s}$) by performing a sum (first row of $H^{-1}$) while the ℓ = 1 signal (second component of $\hat{s}$) is recovered by performing a difference (second row of $H^{-1}$), and this is expressed as a particular case of standard MIMO detection (called linear detection, i.e. inversion of the channel matrix) in the previous equation.

Now we established that the experiment in [1] can rigorously be seen as a MIMO problem, from the discussion above, the second important conclusion of this Comment can be drawn, namely the vorticity of the modified reflector is actually not needed to achieve this result, since the same performance could be achieved with any pair of Tx antennas provided that enough diversity in the linear combination coefficients is available at the two Rx antennas.

## 3. An alternative experiment

To better and intuitively illustrate the above claim, let us consider the alternative experiment symbolically depicted in Fig. 1. We will show that this experiment is similar in essence to the one of [1], further demonstrating that the use of the OAM vortex mode is unnecessary here, and also drawing important conclusions with regard to the applicability of the concept.

Comment on 'Encoding (...) through radio vorticity: first experimental test'

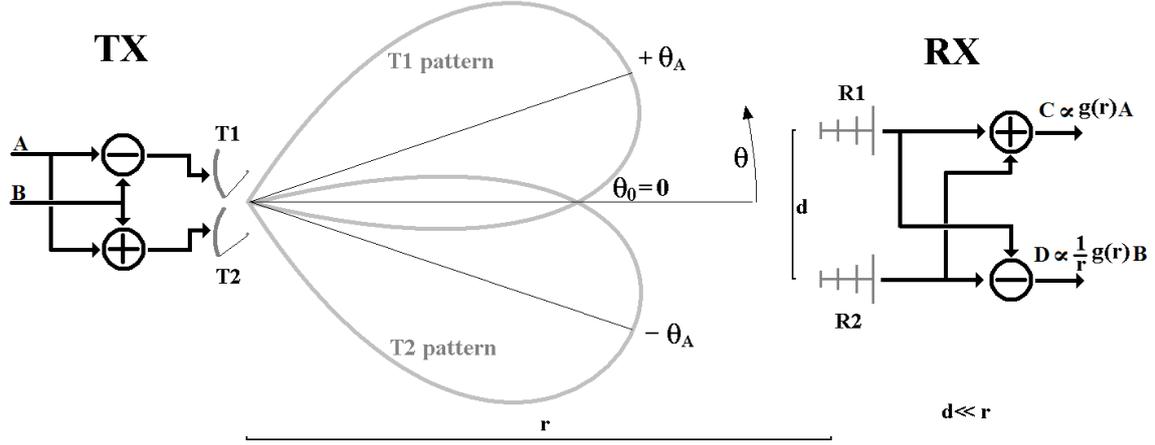

**Figure 1.** Schematic of an alternative experiment leading to equivalent performances without using OAM modes.

In order to allow a more compact form of the equations provided next, we make a number of simple assumptions on the system of Fig. 1, *all without loss of generality.* First, the two Tx directional antennas are the same, are placed close to each other and have approximately the same phase center. They point in two slightly different directions, and we assume that a single polarization is used and that the system is symmetrical with respect to the horizontal axis.

Unlike the Venice experiment, here the two input signals (called A and B) are not fed directly to the Tx antennas, but are treated as a common and a differential mode (left hand side of Fig. 1). In this way, we create the differential mode that will be shown to be similar to the 'vortex' mode of [1] as far as detection is concerned, while using standard antennas. Note also that the two Tx patterns are pointing slightly outwards the line of sight between Tx and Rx centers ($\theta_0 = 0$ in Fig. 1). In this region the patterns $e_{T_i}(\theta)$ vary rapidly with $\theta$, and with opposite slopes for the two patterns, which simply allows better detection sensitivity. The Rx architecture is unchanged: performing the sum of the signals received by the two Rx antennas we can retrieve the signal *A* (that works as a common mode signal) and, provided that the distance *d* between the two Rx antennas is sufficient, the difference will be proportional to signal *B*.

In fact, the electric field (written as a scalar as we consider a single polarization) generated by the two antennas can be approximated by a first order Taylor expansion in the neighbourhood of the Rx node (assuming $d \ll r$ and a radiation pattern without abrupt angular changes as is always the case with finite radiating structures)

$$E_{T1} = g(r)e_{T1}(\theta)(A+B) = g(r)\left[e_{T1}(\theta_0) + \Delta\theta \frac{\partial e_{T1}}{\partial \theta}\right](A+B) \quad (3)$$

$$E_{T2} = g(r)e_{T2}(\theta)(A-B) = g(r)\left[e_{T2}(\theta_0) + \Delta\theta \frac{\partial e_{T2}}{\partial \theta}\right](A-B) \quad (4)$$

where $g(r) = (4\pi r)^{-1} \exp(-jkr)$, $\theta_0 = 0$ is the angle with respect to the axis, *k* is the wavenumber and *A* and *B* are the a-dimensional complex envelopes of the two input signals. Since the structure is symmetrical:

$$e_{T1}(\theta_0) = e_{T2}(\theta_0) \equiv e_0 \quad , \quad \frac{\partial e_{T1}}{\partial \theta} = -\frac{\partial e_{T2}}{\partial \theta} \equiv \partial e \quad (5)$$

Since $d \ll r$, at the two receivers we can set $\Delta\theta = \theta \cong \pm d/2r$, and then evaluate the field:

$$E_{R1} = E_{T1}(+d/2r) + E_{T2}(+d/2r) = g(r)\left[2Ae_0 + 2B\frac{d}{2r}\partial e\right] \quad (6)$$

$$E_{R2} = E_{T1}(-d/2r) + E_{T2}(-d/2r) = g(r)\left[2Ae_0 - 2B\frac{d}{2r}\partial e\right] \quad (7)$$

Comment on 'Encoding (...) through radio vorticity: first experimental test'

And hence (keeping $d$ constant):

$$C \propto E_{R1} + E_{R2} = g(r)4Ae_0 \propto g(r)A \qquad (8)$$

$$D \propto E_{R1} - E_{R2} = g(r)4B\frac{d}{2r}\partial e \propto \frac{g(r)}{r}B \qquad (9)$$

Notice that the common signal $A$ is attenuated as $r^{-1}$, while the differential one is attenuated as $r^{-2}$. As can be intuitively inferred from Figure 1, if $r$ increases, the exploitable 'difference' between the two patterns decreases like $r^{-1}$, assuming that $d$ is kept constant. Having more antennas both at the Tx and Rx nodes, the number of multiplexed signals can be incremented by using more complex interferometers to implement the (pseudo)inverse of H, but only one mode can be common (in the single polarization case): the other ones are differential and rely on the Rx antennas mutual distance $d_{ij}$ and thus also experience at least the same loss factor. This would be the case of any higher OAM mode added to the experiment of [1].

## 4. Discussion

The Rx technique used for separating the two signals is the same in the two experiments. The only difference is the transmitting section, where the differential mode is created by the feed section (Fig. 1) or by the antenna itself [1]. In fact, in both the experiments one mode is received as even (such as the $\ell = 0$ mode) and the other one as odd (such as the $\ell = 1$ mode). When receiving the even mode, the two antennas are working in phase, as a 'normal' broadside array. On the contrary, the odd mode is detected by exploiting a difference of the received field in two different positions (i.e. in two slightly different directions with respect to the Tx array). In both cases an interferometer implements the (pseudo)inverse of $H$, decoding the original signals. Receiving the odd mode is made possible by using two Tx patterns varying with space in different ways: in the experiment of [1] one pattern has a large variation in the receiver area (the vortex antenna) while the other pattern is almost constant (the Yagi-Uda antenna), while in the alternative setup of Fig. 1 we create this variation by placing the receiver where both Tx antenna patterns vary rapidly.

Whichever is the method used to transmit the signals, the patterns are necessarily continuous angular functions since the radiating structures are finite. Therefore, applying the Taylor expansion and following the same derivation as in the alternative example, it is clear that the odd mode signal transmitted by the vortex antenna will also suffer from the additional $r^{-1}$ factor (or $r^{-2}$ in terms of signal power), for a fixed Rx antenna spacing $d$. Importantly, this additional loss factor *cannot* be compensated by trying to modify the Tx antennas orientation or patterns, since the patterns will always be continuous versus $\theta$.

The discussion so far was based on considerations on the fields and antenna patterns, but our conclusions are also in line with well-known MIMO system results (as explained earlier any multiple antenna system can be rigorously seen as MIMO). In particular if we consider a single-polarization, line-of-sight and far-field MIMO system, it is well known [3-5] that the corresponding $H$ matrix will tend to a rank-1 matrix, i.e. only one signal can be practically transmitted. Attempting to transmit two distinct signals would mean not being able, at the Rx node, to demultiplex their superposition. This is due to the fact that the vector of the linear combination coefficients (i.e. a row of $H$) is the same for all Rx antennas, except for a phase factor that, being common to all the vector components, still leads to a rank 1 linear application. If we remove the constraint on the polarization, we can achieve a rank 2 limit $H$ matrix, i.e. 2 signals can be transmitted together (as in TV broadcasting satellites) but not more. This means that in a single polarization, line-of-sight, far-field situation only one of the OAM modes can be successfully transmitted. Since the $\ell = 0$ is a non twisted 'standard' mode, it obviously can be transmitted, and hence it follows that it should not be possible to transmit all the other modes with $\ell \neq 0$ in far-field conditions.

Therefore there is an apparent contradiction between these well-known MIMO results and the capability of decoding the two signals in [1]. This is actually not the case, and the explanation lies in the far-field condition for which the above well-known MIMO properties are derived. In brief, the Venice experiment was not held in a 'far-enough-field' situation: the condition to consider a MIMO system as far-field is: $r \gg$

$2D^2/\lambda$, where $D$ is the size of the larger antenna array (either the Rx or the Tx). The threshold $r_0 = 2D^2/\lambda$ is usually considered as an indicative separation point (see [6]) between the Fresnel near-field and the Fraunhofer far-field, but near-field quantities can still be measured (even if highly attenuating with $r$) also in the "inner" Fraunhofer region. In the experiment of [1] $\lambda = 12.4$ cm, and taking $D = d = 4.5$ m (the Rx antennas separation), we obtain $r_0 = 327$ m. The distance $r$ (442 m) is in the same order of magnitude, and hence the detection of the OAM modes is still possible, which reconciles our analysis with well-known MIMO results. Finally it is noticeable that in the experiment of [1] the scattering from nearby walls, polarization imperfections and the distance between Tx antennas could have aided the detection as well.

## 5. Conclusions

Despite the originality of the idea and the positive outcome of the experiment in [1], in our opinion the authors did not identify that the developed system actually implements a specific type of MIMO system, and in turn that 'vortex' OAM modes are not needed for achieving such a result. In our view, this contradicts the introduction of [1] that suggests that the proposed concept is different from –and overcomes– usual spatial diversity techniques.

Moreover, our analysis reveals that this kind of techniques leads to an additional important received power reduction for a fixed spacing between Rx antennas, which seriously questions the applicability of the concept. If instead $d$ is scaled in order to keep $r/d$ constant, the results are unpractical: if the system used in the Venice experiment is scaled, for example, to satellite distances in the order of $10^3$ Km, then a hypothetical Rx array would have a size $d$ in the order of 10 Km. The alignment would also be very sensitive in the proposed setup, probably preventing the usage for broadcast applications. In other fixed line-of-sight radio links the idea could be used, but alternative MIMO solutions provide similar or better results. Importantly, it should be noted that the beneficial use of vortex modes in guided communication is in no contradiction with our conclusions, since our comment –and the original paper– only relate to the use of OAM modes in wireless systems.